\documentclass[12pt]{article}

\usepackage{a4}
\usepackage{a4wide}

\usepackage{amsmath}
\usepackage{amscd}
\usepackage{amsfonts}
\usepackage{amssymb}
\usepackage{mathrsfs}

\def \be {\begin{equation}}
\def \ee {\end{equation}}
\def \beal#1 {\begin{align}#1\end{align}}

\makeatletter

\@addtoreset{equation}{section}
\makeatother

\begin{document}

\begin{titlepage}
\title{
\vspace{1.5cm}
Stellar Physics and General Relativity
\vspace{1.cm}
}
\author{
Shuichi Yokoyama\thanks{syr18046[at]fc.ritsumei.ac.jp},\; 
\\[25pt] 
${}^{*}$ {\normalsize\it Department of Physical Sciences, College of Science and Engineering,} \\
{\normalsize\it Ritsumeikan University, Shiga 525-8577, Japan}
}

\date{}

\maketitle

\thispagestyle{empty}

\begin{abstract}
\vspace{0.3cm}
\normalsize
The general theory of relativity is currently established as the most precise theory of gravity supported by observations, and its application is diverse ranging from astronomy to cosmology, while its application to astrophysics has been restricted only to compact stars due to the assumption that the Newtonian approximation is sufficient for celestial bodies with medium density such as the sun. Surprisingly, the recent research of the author has implied that this long-held assumption is not valid, and that non-perturbative effects significantly change relevant results obtained by Newtonian gravity. In particular, local physical quantities inside the sun are newly predicted to exhibit power law differently from the so-called standard solar model. This surprising result is reviewed including brief discussion of physics behind the discrepancy and a new application of the new mass formula to gas planets. 

\end{abstract}
\end{titlepage}
%\tableofcontents

\maketitle

\section{Introduction}

The general theory of relativity (GR), which was invented by Einstein's original top-down approach, has elucidated and predicted new gravitational phenomena that could not be explained by the Newton's theory of gravity, and has currently established its status as the most precise theory of gravity supported by observation. The application of GR is wide from astronomy to cosmology and has been verified until these days.

However, the application of GR to astrophysics has been limited only to celestial bodies with high density, such as neutron stars. This is because it has been believed that the Newtonian approximation is sufficient for astronomical bodies with medium density such as main sequence stars including the sun. 
Conventionally, such astronomical bodies with medium density have been investigated by using the so-called stellar structure equations described already in an old textbook based on Newtonian gravity \cite{eddington1930internal}, and, in particular, physical observables inside the sun such as sound speed and density were computed by using the classic structure equations. (See \cite{RevModPhys.60.297,carroll2007introduction}, for instance.) These results were tested by solar seismic data \cite{1991sia..book..401C,ca43c68c-25f4-31cc-8684-6acb8f9f00b7}, and the results are collectively known as the standard solar model. (See \cite{Antia2003,Christensen_Dalsgaard_2021} for instance.)

The author has been working on non-perturbative aspects of theoretical physics mainly to challenge the issue of quantum theory of gravity. During the work, he realized that the conventional definitions of energy in curved space-time such as the quasi-local energy and the Komer mass are not precise or strictly not correct in general, and proposed its definition to improve it \cite{Aoki:2020prb}. It turned out that the proposed definition was already written in an old textbook \cite{Fock:1959}, but its validity and usefulness were demonstrated first by proving that the mass of classic black holes were reproduced from the definition.
While the proposed definition incorporates the relation between symmetry and conservation elucidated by Noether, it is easily seen therefrom that there exists a new conserved charge not necessarily associated with symmetry. 
The non-Noether conserved charge was proposed to describe the entropy of the system and some evidence was provided by reproducing entropy for the homogeneous isotropic universe and for classic black holes \cite{Aoki:2020nzm}. 
The proposed entropy was subsequently computed for the system of an arbitrary hydrostatic equilibrium system with spherical symmetry, and it was shown that the proposed entropy density satisfies the local Euler's relation and the first law of thermodynamics concurrently and non-perturbatively in the Newton constant \cite{Yokoyama:2023nld}, and that the thermodynamic observables are analytically determined for a hydrostatic equilibrium system with uniform density or incompressibility. (See also \cite{Oppenheim_2003}.)
This result implies that a set of differential equations fully determine the relativistic internal structure of a star with an equation of state given, and the author has proposed the set of the differential equations as a relativistic extension of the classic stellar structure equations \cite{Yokoyama:2023gxf}. 

The purpose of this piece of proceedings is to review the proposed relativistic structure equations and their relevant consequences and predictions including a little new application of the new mass formula to gas (dwarf) planets. 

\section{Relativistic hydrostatic structure equations}
\label{RHSE}

The proposed relativistic hydrostatic structure equations are given as follows \cite{Yokoyama:2023gxf}.
\beal{
\frac{dP}{dr} =& -\frac{(P+\rho)}{c^2} \frac{ G( \check E_r+4\pi r^3 P)}{ r^2(1-2\frac{ G\check E_r }{rc^2}) }, 
\label{TOV} \\
\frac{d\check E_r}{dr}=& 4\pi r^2\rho,
\label{rho} \\ 
\frac{dT}{dr} =& - \frac{ G( \check E_r+4\pi r^3 P)}{c^2 r^2(1-2\frac{ G\check E_r }{rc^2}) } T. 
\label{T}
}
The 1st and the 2nd equations are already known in \cite{PhysRev.55.374}. The 1st one describes the balance between the attractive force of gravity and the repulsive force of matter in relativistic hydrostatic equilibrium with spherical symmetry. Indeed, in the Newtonian approximation such that the contribution of pressure $P$ is much smaller than that of energy $\check E_r$, $4\pi r^3 P \ll \check E_r$, and that of energy density $\rho$, $P\ll\rho$, and that the region is far away from the horizon radius of the black hole with mass $\check E_r/c^2$, $2\frac{ G\check E_r }{c^2}\ll r$, the 1st equation \eqref{TOV} reduces to the non-relativistic hydrostatic equation 
\be 
\frac{dP}{dr} = -\frac{ G_{\rm N}\varrho M_r}{ r^2 },
\label{NRHydrostatic}
\ee
where $M_r = \check E_r/c^2$, $\varrho=\rho/c^2$ is the mass density, and $G_N=Gc^2$ is the Newton constant. 

The 3rd equation is required for the system to be in local thermodynamic equilibrium. It is well-known in thermodynamics and statistical physics that there is an identity among thermodynamic quantities, one of which is $T(\frac{\partial P}{\partial T})_{V,N} = (\frac{\partial U}{\partial V})_{T,N} +P$. This identity held in global thermodynamics is expected to hold also in local thermodynamic equilibrium. That is, in a local thermodynamic equilibrium system with spherical symmetry, it must satisfy
\beal{
T\frac{dP}{dr} = (\rho +P)\frac{dT}{dr}.
\label{LTE}
}
The 3rd equation \eqref{T} is obtained by substituting \eqref{TOV} into \eqref{LTE}.
The relativistic structure of a spherically symmetric hydrostatic equilibrium system whose state is specified by pressure, energy density, and temperature is determined by solving the above three differential equation with the addition of one equation of state such that $P=P(\rho,T)$. 

There are some comments. 
It will be soon realized that there is no luminosity variable in the proposed relativistic structure equations. This is a natural consequence of solving the Einstein equation with a perfect fluid and actually preferable in the sense to describe a wide class of astronomical bodies including non-luminous stellar objects. 
It has been claimed that the proposed relativistic structure equations are applicable to luminous stars with luminosity sufficiently small or, more precisely, much smaller than a certain bound corresponding to the Eddington bound such that $L_r \ll \gamma \frac{4\pi cG( \check E_r+4\pi r^3 P)}{\kappa (1-2\frac{ G\check E_r }{rc^2})^{\frac32} }$, where $\gamma$ is the heat capacity ratio of the stellar fluid and $\kappa$ is the stellar opacity.
The luminousity variable in the stellar structure equations can be included by incorporating radiative fluid though its inclusion has to be treated separately from main stellar fluid due to its null property. 

Since the system is in hydrostatic equilibrium, thermodynamic observables used in the structure equations are expected to satisfy thermodynamic relations. Indeed, it was shown in \cite{Yokoyama:2023nld} that the entropy density for this system can be  constructed as the non-Noether charge following the proposal in \cite{Aoki:2020nzm}, and that the entropy density together with the thermodynamic quantities in the sturcture equations satisfies both the local Euler's relation and the first law of thermodynamics non-perturbatively with respect to the Newton constant, where the local temperature exactly coincides with the Tolman temperature related to the gravitational potential $\phi$ as $T \propto e^{-\phi/c^2}$ \cite{PhysRev.35.904}. (See also \cite{landau}.) 

The proposed relativistic structure equations must be consistent with the classic ones except for the equations containing the luminosity variable in the Newtonian regime. That is, the equation \eqref{T} must reduce to the one which is supposed to hold in the convection zone 
\beal{
\frac{dT}{dr}= -(1 - \frac1\gamma )\frac{G_{\rm N} M_r \varrho }{r^2P}T,
\label{Convection}
}
This was done by applying the structure equations to a system of ideal gas with its particle number conserved, as explained in the next section \ref{Applications}. 

\subsection{Exact relativistic Poisson equation and steady-state heat conduction equation}
\label{PoissonEquation}

Before proceeding to the reproduction of \eqref{Convection}, it is convenient to see some exact results derived only from the proposed stellar structure equations without using an equation of state, which is also helpful to check the validity thereof. 
To the end, first rewrite the 3rd equation \eqref{T} by using the gravitational potential as 
\beal{
\check E_r=\frac{r^2 \phi'/G-4 \pi  r^3 P}{1+ 2r \phi'/c^2}, 
}
Differentiating both sides and substituting \eqref{rho}, the equation can be rewritten in the following form
\beal{ 
\nabla^2 \phi =4 \pi  G_{\rm N} (\rho + 3 P )/c^2, 
\label{GRPoisson}
}
where the covariant derivative is defined on the spherically symmetric metric $g_{\mu\nu}\mathrm dx^\mu \mathrm dx^\nu
=- e^{2\phi/c^2} \mathrm (cdt)^2 +g_{rr} \mathrm dr^2 + r^2(\mathrm d\theta^2+(\cos\theta)^2\mathrm d\phi^2)$. 
This form of the relativistic Poisson equation in the curved spacetime had been derived in the weak gravity approximation \cite{zel2014stars}, but the form \eqref{GRPoisson} is in fact exact, which was first proved in \cite{Yokoyama:2023gxf} as far as the author knows.
Notice that the relativistic Poisson equation is non-linear in general curved spacetime. 

On the other hand, the relativistic Poisson equation can be rewritten in terms of the local temperature by using the Tolman relation as 
\beal{ 
\nabla^2( \log T )= - 4 \pi  G (\rho + 3 P )/c^2.
\label{HCE}
}
This is the steady-state heat conduction equation in relativistic hydrostatic equilibrium, which determines the temperature profile inside a celestial body including the full relativistic effect. 

It is important to stress that these results \eqref{GRPoisson}, \eqref{HCE} are derived exactly. The importance of exactness is seen in the next section of application to the construction of a relativistic stellar model.

\section{Applications}
\label{Applications}

\subsection{Ideal gas of baryonic particles}
\label{IdealGas}

For the purpose to reproducing the structure equation in the convection zone \eqref{Convection} and investigating the relativistic effect in stellar physics, consider a hydrostatic equilibrium system of an ideal gas of particles with their number current conserved, which are called baryonic particles for convenience. 
The equation of state for an ideal gas is $P = \hat n k_{\rm B} T,$ where $\hat n$ is the number density. 
The conservation of the particle number current imposes the number density to satisfy $\frac{d\hat n}{dr} = \frac {\hat n}{\rho + P}\frac{d\rho}{dr}$.
Combining these equations of state, one can solve the differential equation to describe the local thermodynamic identity \eqref{LTE} as $P=w\rho$, where $w$ is an integration constant. It can be fixed by using the expression of the energy density $\rho =\hat nC_VT$, where $C_V$ is the heat capacity with fixed volume as $w=\frac{k_{\text B}}{C_V} = \gamma - 1$, where $\gamma$ is the heat capacity ratio. Plugging this back into \eqref{LTE} leads to 
\be 
\frac{dT}{dr} =(1-\frac 1{\gamma})\frac TP\frac{dP}{dr}.
\label{SSC}
\ee
This is in fact the saturation point of the inequality of the Schwarzschild stability for hydrostatic equilibrium \cite{1906Schwarzschild}. By substituting the non-relativistic from of the hydrostatic equation \eqref{NRHydrostatic} into this, the structure equation in the convection zone \eqref{Convection} is reproduced. 
Note that the equation of state of the form $P=(\gamma - 1)\rho$ is necessary and sufficient to obtain \eqref{SSC} from \eqref{LTE}. 

This new derivation of the equation of convective state \eqref{SSC} elucidates that an ideal gas of baryonic particles is in local thermodynamic equilibrium at the saturation point of the inequality of the Schwarzschild stability condition, while there is another immediate logical consequence that the heat capacity ratio is approximately one in the convection zone with the Newtonian approximation valid. 
This is a simple prediction and is expected to be testable underneath the solar surface. 

Another important consequence is that the steady-state heat conduction equation can be solved exactly in the equation of state.
To see this, first solve \eqref{SSC} to determine pressure as $P\propto T^{\frac\gamma{\gamma-1}}$, and thus the energy density is fixed as $\rho=aT^{\frac\gamma{\gamma-1}}$, where $a$ is a constant. Substituting these into \eqref{HCE}, the steady-state heat conduction equation is written only in terms of temperature variable, where each term consists of its power. This suggests the existence of a power law solution, and indeed such a solution exists as 
\be 
T = \left(\frac{c^2 (\gamma-1)}{2 \pi  a G \left(\gamma^2+4\gamma -4\right)}\right)^{1 - \frac{1}{\gamma}} \frac1{r^{2(1 - \frac{1}{\gamma}) } }.
\label{Tsol}
\ee
Accordingly, pressure and energy density similarly obeys the power law as $P\propto \rho\propto 1/r^2$. 
Remark that the Newton constant appears in an inverse power. This implies that this solution cannot be reach perturbatively by post-Newtonian analysis.
The exact analysis of the relativistic extension of the stellar structure equations uncovers the power law behavior of thermodynamic observables in stellar physics. 

\subsection{Comment on application to corona}
\label{corona}

The power law behavior of the temperature implies the applicability of the above result to a region in plasma state such as corona, since it is known in plasma physics that temperature in a system of ionized ideal gas falls off in power law. (See \cite{choudhuri_1998} for instance.) 

In the past, corona was investigated by Parker combining this result of plasma physics with the non-relativistic hydrostatic equation \eqref{NRHydrostatic} \cite{1958ApJ...128..664P}. He pointed out that pressure of the system is non-vanishing at far asymptotic region, and argued that this was a signal of gas in solar corona to be out of hydrostatic equilibrium resulting with the solar wind, whose existence had been suggested by \cite{1951ZA.....29..274B}. 

However, the above result of the relativistic hydrostatic equations and this conventional one are inconsistent because pressure in the former also falls off in a power law. The author has pointed out a flaw in the conventional argument that the Newtonian approximation breaks down in the latter solution with pressure term in the hydrostatic equation non-negligible. 

\subsection{New stellar model}
 
A new stellar model has been constructed by combining the relativistic hydrostatic structure equations explained in section \ref{RHSE} and ideal gas of baryonic particles in \ref{IdealGas}. Here, for simplicity, review briefly some relevant points of the model simplifying the situation. See \cite{Yokoyama:2023gxf} for more detail. 

As explained in section \ref{IdealGas}, the internal macroscopic observables obey power law as 
\be 
\rho= \rho(R) (\frac{R}{r})^2, \quad 
P= (\gamma-1)\rho, \quad 
T= T(R)(\frac{R}{r})^{2(1 - \frac{1}{\gamma}) }, 
\label{PowerLaw}
\ee
where $R$ is the radius of the stellar surface.
These are determined once two inputs at boundary, the surface density $\rho(R)$ and the surface temperature $T(R)$, and the heat capacity ratio $\gamma$ are fixed. 
Note that a singularity develops at the center for these quantities. Therefore the power law behavior must change near the center, whose region is called the core. 

There is an immediate consequence of the power law behavior \eqref{PowerLaw} about how to estimate the mass of a star.
Using \eqref{PowerLaw}, the 2nd structure equation \eqref{rho} can be easily integrated to $\check E_r = \rho(R) 4\pi R^2 r$. Thus the mass of the star $M$ is evaluated as 
\be 
M = \check E_R/c^2 = \varrho(R) 4\pi R^3, 
\label{MassFormula}
\ee 
where $\varrho(R)=\rho(R)/c^2$ is the mass density at the stellar surface. 
This implies that the stellar mass can be estimated by its surface density multiplied by its volume up to numerical factor as long as the stellar interior is mainly constituted by the ideal gas of baryonic particles. 
To see the validity, apply this formula to the sun, which mainly consists of hydrogen, so that its surface density may be approximated as $\varrho_{\odot} \approx \chi m_{\rm H}/(\frac43\pi a_{\rm B}^3)$, where $\chi$ is the filling factor at the solar surface, $m_{\rm H}$ is the mass of hydrogen and $a_{\rm B}$ is the Bohr radius. Then the mass formula for the sun leads to $M_\odot \approx 3\chi m_{\rm H} (R_\odot/r_{\rm B})^3$, where $R_\odot =7\times 10^{8}{\rm m}$ is the solar radius. 
Choosing the filling fraction as $\chi=1/6$ by hand, the solar mass is evaluated as $M_\odot\approx 2\times10^{33}{\rm g}$, which reproduces observational data \cite{doi:10.1142/11218}.

The result of the application of the mass formula \eqref{MassFormula} to other gas (dwarf) planets on the assumption of the same surface density as the sun is presented in Table~\ref{MassData}. 
\begin{table*}[t]
 \caption{ The table shows the comparison of observational data of the mass for gas (dwarf) planets and its predicted value by the mass formula \eqref{MassFormula} on the assumption of the same surface density as the sun. The mass formula is seen to work at least as order estimate without detailed data for each star, and particularly a good agreement is seen for Jupiter, which mainly consists of hydrogen as the sun.
}
 \begin{center}
  \begin{tabular}{|c|cc|c|}
  \hline
Gas planets  & Radius ratio $(R/R_\odot)$ & Mass ratio $(M/M_\odot)$  & Mass formula $(R/R_\odot)^3$ \\
 &\multicolumn{2}{c|}{ Observation } & Prediction \\ 
   \hline
Jupiter &  $1.0 \times 10^{-1}$     & $9.6\times 10^{-4}$ & $1.0 \times 10^{-3}$ \\
Saturn &  $8.7 \times 10^{-2}$     & $2.9\times 10^{-4}$ & $6.5\times 10^{-4}$ \\
Uranus &  $2.6 \times 10^{-2}$     & $4.4\times 10^{-5}$ & $5.0\times 10^{-5}$ \\
Neptune &  $2.5 \times 10^{-2}$     & $5.2\times 10^{-5}$ & $4.5\times 10^{-5}$ \\
Pluto &  $1.7 \times 10^{-3}$     & $7.4\times 10^{-9}$ & $5.1\times 10^{-9}$ \\
    \hline
  \end{tabular}
 \end{center}
 \label{MassData}
\end{table*}
The reason for this assumption is based on the nature of particles that light comes up and heavy goes down, so that gas near the surface is expected to consist of light components such as hydrogen. 
The validity of the formula with the assumption is tested by the agreement between the observation of the mass and its prediction by the mass formula, and it is seen at least in the order level even though their internal constituents are different to each other. In particular, the agreement for the case of Jupiter is quite good, which is reasonable because its main constituent is hydrogen similarly to the sun. 
This result suggests that this mass formula can be also used as a standard to guess the content of the stellar internal constituent. 

To fix the temperature profile inside the sun, use the solar surface temperature as $T(R_\odot)=T_\odot \approx 6000{\rm K}$, and assume that the heat capacity ratio at the surface is $\gamma(R_\odot)=1$, as argued in section \ref{IdealGas}. In addition, use an observational information that the convective region transits to the radiative one at solar tachocline existing around $r\approx 2/3R_\odot$. Then, $\gamma(2/3R_\odot)=4/3$, since the radiation zone starts at this point. 
Furthermore, it needs to fix relevant properties of the core. 
Here assume that the core radius is $r_{\rm c}=5r_{\rm H}$, where $r_{\rm H}=2M_\odot G_N/c^2 \approx 3{\rm km}$, and that it is supported by repulsive force of degenerate electron gas as a white dwarf \cite{10.1093/mnras/87.2.114}, so that the heat capacity ratio is fixed as $\gamma(r_{\rm c})=5/3$ \cite{prialnik2000introduction}. 
In \cite{Yokoyama:2023gxf}, the heat capacity ratio in the intermediate regions was approximately fixed by the linear interpolation for convenience, though it is something to be fixed by observation.   

Using these inputs, thermodynamic observables are easily computed without complicated numerical calculation. In particular, at the core, temperature is computed as $T(R_{\rm c})= T(R)(R/R_{\rm c})^{5/4}\approx 2\times 10^7{\rm K}$, and the density is $\rho(R_{\rm c})/c^2=\rho(R)/c^2(R/R_{\rm c})^2 \approx 2\times 10^{11} {\rm kg/m^3}$. 
The core temperature in this new solar model is consistent with that of the standard one \cite{RevModPhys.60.297,carroll2007introduction}, so that it increases highly enough to ignite self-sustaining nuclear chain reaction such as PP-chain and CNO cycle. On the other hand, the core density is much bigger than the one of the standard solar model, $1.5 \times 10^{5} {\rm kg/m^3}$, and in fact it is bigger than the average density of a typical white dwarf, $3\times10^9 {\rm kg/m^3}$, but smaller than that of a neutron star around $5\times10^{14} {\rm kg/m^3}$ \cite{glendenning2012compact}. 
This high value of the core density, which is reasonable due to the assumption for the core to be supported by degenerate electron gas, implies a new scenario of the evolutionary process of a luminous star which is supposed to evolve into a white dwarf via a red giant and a planetary nebla by blowing outer layer off gradually. In the standard solar model, during the last process to blow internal matter off, a certain gravitational collapse concurrently happens so as to increase the density up to that of a white dwarf, while in the new solar one, such a gravitational collapse does not need to happen: the core remains more or less unchanged, or similarly blow off some portion. 

The profiles of thermodynamic variables were plotted in \cite{Yokoyama:2023gxf}. Their behavior is qualitatively different from the conventional result known as the standard solar model. 
In the new solar model, all thermodynamic observables increase rapidly only near the core due to their power law, while, in the standard one, temperature grows almost linearly from the surface and density comparatively spreads away from the core. 
While the discrepancy of temperature starts almost from the neighborhood of the surface, that of density is seen in a deep interior region with $r \lesssim 0.3R_\odot$. 

\section{Discussion}
\label{Discussion}

A relativistic extension of the classic stellar structure equations has been proposed. The set is given by the TOV equation with the known mass gradient equation and the local version of a thermodynamic identity. From the proposed relativistic structure equations, the exact forms of the relativistic Poisson equation and steady-state heat conduction equation have been derived, which are converted from one to the other by the Tolman relation of the equivalence between gravitational potential and local temperature. 
The application to an ideal gas of baryonic particles has been reviewed, which provides the necessary and sufficient condition to reproduce the differential equation in convection zone as well as plays a role of the main content to construct a new stellar model. In the system, the steady-state heat conduction equation has been solved exactly, and as a non-perturbative result, the thermodynamic observables obey power law. 
A new stellar model using this ideal gas has been reviewed. It provides a new mass formula, which works well for several gas planets. Accordingly a new solar model is also presented. It has turned out that the predicted power-law behavior of the macroscopic observables by the new solar model is qualitatively different from that of the standard one. 

The discrepancy between the presented new solar model and the standard one is non-negligible. 
Then a question is, which is selected by nature. A conventional or conservative physicist might support the standard solar model because, as mentioned in introduction, results of sound speed and density in the standard solar model were already tested by helioseismology, and their agreement looks so good that there is no room for new physics as the name suggests. 

However, there are several unreasonable or unsatisfactory points in the standard solar model. 
On density, as pointed out, its profile in the standard one spreads non-trivially away from the core. This result physically means that the repulsive force near the core becomes so strong as to push matter away. Then the pressure term in the hydrostatic equation \eqref{TOV} would not be negligible. Indeed, the region near the core is supposed to be dominant by radiation, and, if it is the case, then the pressure term certainly could not be neglected since radiant pressure is known to be comparable to the internal energy density as known in statistical physics. There could be possibly a similar flaw in the standard argument of the solar interior as pointed out that in corona in section \ref{corona}. In addition, it is known in community that raw seismic data are contaminated by sizable background noise, and the removal thereof to extract physics requires certain expertise. It is not clarified at least to the author why or whether the conventional result of helioseismology is not affected by such background noise in the deep inside of the sun. 
It may be worth scrutinizing the result of density 
by helioseismology. 
On temperature, its discrepancy is more serious. In the new solar model, the high critical temperature of nuclear chain reaction is achieved only near the core, while it is in a middle depth in the standard one. In the latter, the nuclear chain reaction is supposed to occur in a vast region of the sun. Then a question arises that there might cause a trouble to use up hydrogen before the solar life time. 
It also seems that there is no strong support for the latter result from seismic data so far. 

The author is asking which nature selects and you bet. 

\section*{Acknowledgement}
This piece of proceedings is written based on two talks given in IAU General Assembly 2024 held in Capetown, South Africa and IWARA2024 in Cusco, Peru and on a poster presentation in ESPM-17 in Torino, Italy. 
The author would like to thank their participants for discussion. 
This work is supported in part by the Grant-in-Aid of the Japanese Ministry of Education, Sciences and Technology, Sports and Culture (MEXT) for Scientific Research (No.~JP22K03596). 

\bibliographystyle{utphys}
\bibliography{stst3}%

\end{document}